%% file: manuscript.tex
\documentclass[preprint,12pt]{elsarticle}

\usepackage{amssymb}
\usepackage{amsmath}
\usepackage[overload]{empheq}
\usepackage{caption}
\usepackage{subcaption}
\usepackage{array}
\usepackage{bm}

\usepackage{algorithm}
\usepackage{algorithmic}
\usepackage{hyperref}
\journal{Arxiv}

\begin{document}

\begin{frontmatter}



\title{DeepF-fNet: A Physics-Informed Neural Network for Vibration Isolation Optimization}


\author{A. Tollardo, F. Cadini, M.Giglio, L. Lomazzi} %% Author name

\affiliation{organization={Department of Mechanical Engineering, Politecnico di Milano},%Department and Organization
            addressline={Via La Masa n.1}, 
            city={Milano},
            %postcode={}, 
            %state={},
            country={Italy}}

%% Abstract
\begin{abstract}
Structural optimization is essential for designing safe, efficient, and durable components with minimal material usage. Traditional methods for vibration control often rely on active systems to mitigate unpredictable vibrations, which may lead to resonance and potential structural failure. However, these methods face significant challenges when addressing the nonlinear inverse eigenvalue problems required for optimizing structures subjected to a wide range of frequencies. As a result, no existing approach has effectively addressed the need for real-time vibration suppression within this context, particularly in high-performance environments such as automotive noise, vibration and harshness, where computational efficiency is crucial.

This study introduces DeepF-fNet, a novel neural network framework designed to replace traditional active systems in vibration-based structural optimization. Leveraging DeepONets within the context of physics-informed neural networks, DeepF-fNet integrates both data and the governing physical laws. This enables rapid identification of optimal parameters to suppress critical vibrations at specific frequencies, offering a more efficient and real-time alternative to conventional methods.

The proposed framework is validated through a case study involving a locally resonant metamaterial used to isolate structures from user-defined frequency ranges. The results demonstrate that DeepF-fNet outperforms traditional genetic algorithms in terms of computational speed while achieving comparable results, making it a promising tool for vibration-sensitive applications. By replacing active systems with machine learning techniques, DeepF-fNet paves the way for more efficient and cost-effective structural optimization in real-world scenarios.
\end{abstract}

%\includegraphics[width=1\textwidth]{grabs}

%\item A double-network PINN framework for eigenfrequency-oriented optimization is presented
%\item One network solves the inverse eigenvalue problem, the second one enforces physics to improve training
%\item A proposed algorithm facilitates the deployment of the trained DeepF-fNet model
%\item A semi-active locally resonant metamaterial is used to filter user-defined frequencies
%\item DeepF-fNet outperforms genetic algorithms in complex optimization problems

%% Keywords

\end{frontmatter}


\input{sections/introduction}

\input{sections/methodology}

\input{sections/casestudy}

\input{sections/conclusions}

\bibliographystyle{elsarticle-num} 
\bibliography{Thesis_bibliography}

\end{document}

%% file: sections/introduction.tex
\section{Introduction}
\label{sec:introduction}%

Vibration isolation is a crucial aspect of mechanical engineering in various sectors, including automotive, civil, and aerospace. Uncontrolled vibrations can cause adverse acoustic disturbances and even accelerate structural deterioration, requiring effective mitigation strategies.

The most widely adopted countermeasures mainly involve passive vibration isolation systems, with active systems being utilized to a lesser extent \cite{why_active,GARDONIO2000483,LIU200521}. The former method relies on either the design of the structure itself or additional isolating devices to block precise exciting frequencies, while the latter uses actuators to control vibrations. Although relatively easy to design and implement, passive isolators are usually inefficient at low frequencies and when addressing broad-band vibrations \cite{why_active}. Active systems overcome these limitations, but typically involve higher costs and power demands to generate the dynamic force needed to counteract the excitation. In addition, they are more complex due to the incorporation of a control loop, which must be carefully designed for effective operation \cite{LIU200521}.

Building upon passive and active systems, a third approach, termed semi-active vibration isolators, has been introduced as an alternative to fully active control for certain applications \cite{karnopp1974vibration,crosby1973active}. Semi-active control strategies combine the reliability of passive devices with minimal energy consumption, while offering the versatility and performance of fully active systems \cite{LIU200521}. These devices dynamically tune their parameters to adjust their natural frequency spectrum and filter harmful user-defined frequencies. This process may be conceptualized as an inverse eigenvalue problem \cite{doi:10.1137/S0036144596303984,inverseP}, which is characteristically ill-posed and requires additional constraints to achieve a feasible solution \cite{mishra2023estimatesgeneralizationerrorphysics}. Consequently, continuous optimization of the structural parameters is necessary at each sampling period defined by the control system. However, since eigenfrequency-oriented optimization is generally non-linear, it poses significant challenges for existing numerical algorithms.

To address these complexities, genetic algorithms (GAs), inspired by evolutionary biology \cite{TopoReview}, have been widely used for their versatility and robustness. For example, Abdeljaber et al. \cite{ABDELJABER201650} optimized the inner structure of a rod using finite element (FE) simulations to suppress user-defined frequencies. Madeira et al. \cite{inproceedings} employed a GA as the central component of a topology optimization framework to increase the first and second eigenfrequencies of a plate. Although GAs provide accurate solutions to a wide range of problems, they are often time-consuming since their peculiar nature-inspired working principle relies on stochastic improvements of the objective function.

To provide enhanced computational efficiency and performance, more recently, artificial neural networks (ANNs) have been increasingly adopted for their ability to approximate any function \cite{HORNIK1989359}. Among these, physics-informed neural networks (PINNs) have garnered attention for their resource efficiency and superior inference capabilities compared to purely data-driven ANNs \cite{RAISSI2019686,HAGHIGHAT2021113741,JAGTAP2022111402,HU2024127240}. These networks leverage knowledge of the physical laws and constraints of the problem within the total loss function, thus reducing the number of samples required to train the model and improving the generalization capabilities. Although they have been shown to be accurate on the problems on which they are trained, a slight change in the setup necessitates the model to be trained again, which can be a time-consuming process. Building upon the foundation provided by PINNs, deep operator networks (DeepONets) offer an advanced approach by accurately learning the underlying differential operators associated with eigenfrequency problems. This allows for enhanced tolerance to variable inputs \cite{Lu_2021,HE2024117130}, significantly expanding the potential of neural network models to address diverse and dynamic conditions compared to conventional methods. Furthermore, as mathematical surrogates, DeepONets enhance computational efficiency compared to pure physical models, making them well suited for rapid-response optimization in complex problems. The DeepONet framework has been modified by Molinaro et al. \cite{molinaro2023neuralinverseoperatorssolving} through the development of the neural inverse operator (NIO). Once trained, this network is capable of identifying the system parameters of a one-dimensional harmonic wave equation, accommodating variations in boundary conditions within defined limits. However, the eigenfrequency under consideration is fixed, which limits the effectiveness of an algorithm for semi-active vibration isolators. In fact, for optimal performance, such algorithms must operate over a broad frequency range, thus precluding the use of NIOs for this application.

To the best of our knowledge, no existing methods based on DeepONets have been proposed for use in semiactive vibration isolators. To address this gap, the objective of this work is to develop an efficient and reliable model to estimate optimal structural parameters to target a specific unwanted frequency. To rapidly identify optimal structural parameters across an extensive frequency spectrum, we introduce DeepF-fNet, an innovative solution that integrates a dual-network PINN within the proprietary SICE4 deployment algorithm. The method was validated against current state-of-the-art algorithms such as GA to demonstrate the superior performance of SICE4 in terms of computational speed. This algorithm can be adopted in fields where a fast response of the noise isolation device is crucial, such as automotive noise, vibration, and harshness (NVH) \cite{SOLEIMANIAN2024109861}, acting as a potential semi-active replacement for fully active systems.

The remainder of this paper is organized as follows. Section~\ref{sec:methodology} describes the proposed methodology, detailing the components of the SICE4 deployment algorithm and the DeepF-fNet framework. Section~\ref{sec:case study} presents a case study in which a locally resonant metamaterial forms the basis for the testing and validation of SICE4, with its performance compared to that of a conventional GA. Finally, section~
\ref{sec:conclusions} draws out the conclusions from this work and suggests future developments.

%% file: sections/methodology.tex
\section{Methodology}
\label{sec:methodology}%
The proposed framework employs proprietary algorithms, SICE4 and DeepF-fNet, to determine the optimal structural parameters of a metamaterial that allow filtering out a user-specified target frequency.

\subsection{SICE4 Algorithm}
\label{subsec:SICE4}%
The SICE4 deployment algorithm is made up of four steps:
\begin{enumerate}
    \item \emph{System input}: The algorithm is provided with the desired target frequency $\hat{f}$ to filter out.
    \item \emph{Initialization}: The input data for DeepF-fNet consist of $N$ dispersion curves, denoted as $f_{n} = f_{n}(\kappa)$, where $f_{n}$ signifies the $n^{th}$ eigenfrequency and $\kappa$ represents the wave number. Bandgaps, defined as frequency ranges in which there are no values of $f_{n}$ for any $\kappa$, are crucial to determining the filtering effect against external excitations (see Fig. \ref{fig:dispersion_curves}). However, the association between bandgaps and the desired target frequency to filter is inherently complex, necessitating an initial estimation of the curves $f_{n}(\kappa)$. The initialization process involves selecting the initial dispersion curves from a dataset of established spectra. Identification of a candidate spectrum is achieved by minimizing the mean squared error between the central frequency of its initial bandgap and the target frequency, $\hat{f}$. The deviation between these frequencies is quantified as the shift parameter $\Delta$.
    \begin{figure}[h!]
    \centering
    \includegraphics[width=0.7\linewidth]{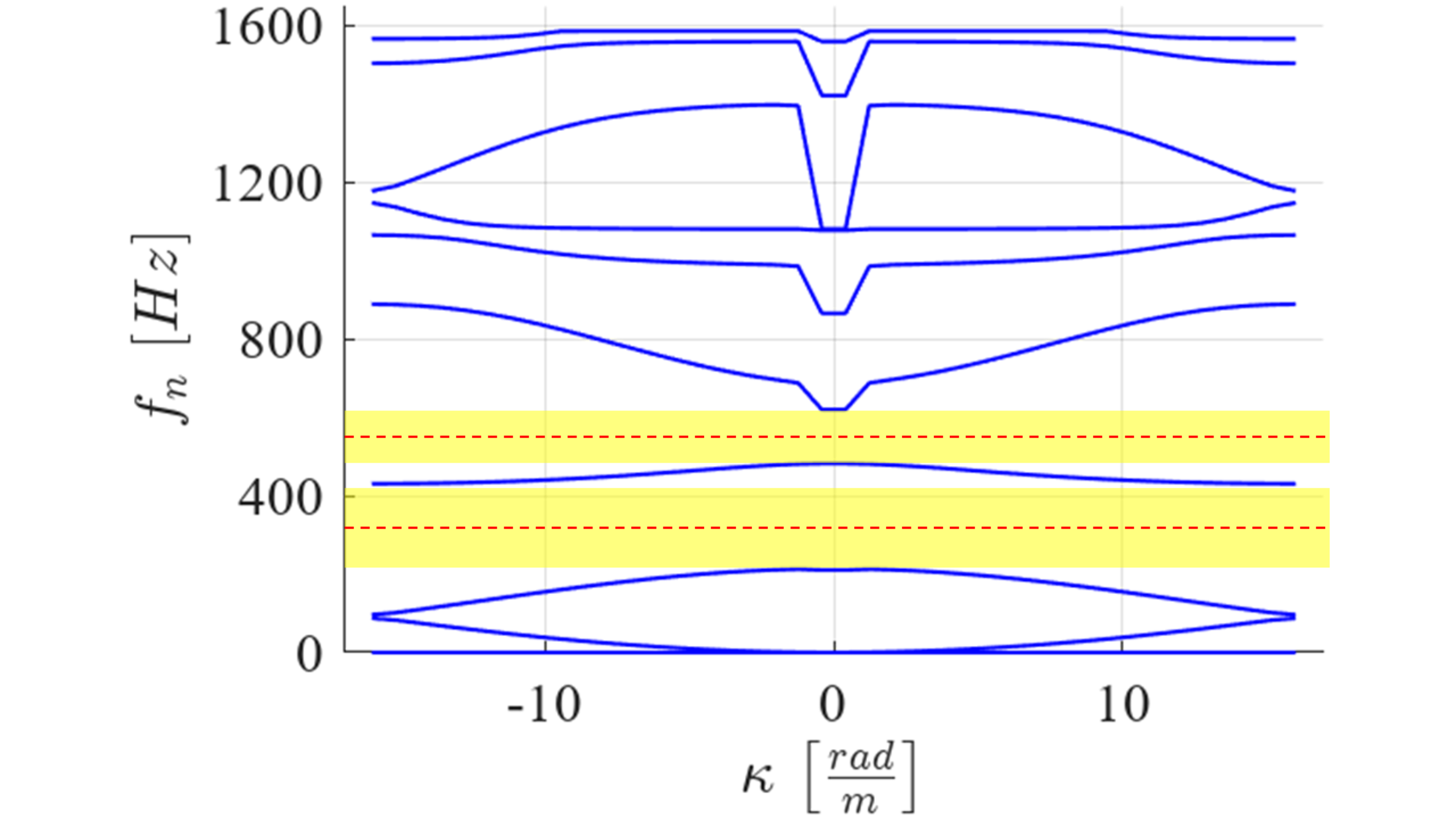}
    \caption{Example of dispersion curves (blue solid lines), with the highlighted bandgaps (yellow) and the relative average frequencies (red dashed lines).}
    \label{fig:dispersion_curves}
    \end{figure}
    \item \emph{Correction}: The parameter \(\Delta\) is applied to shift all dispersion curves in the chosen sample, with the exception of the first, which must be excluded to avoid creating a gap at 0 Hz and to preserve rigid body modes. By so doing, the newly established initial bandgap is more uniformly distributed around the target frequency. Then, to ensure the physical accuracy of the bandgaps, a second correction factor, $\Gamma$, is introduced. It is defined as the ratio between the minimum value of the second dispersion curve and the maximum value of the first dispersion curve within the selected spectrum. The first dispersion curve is then scaled by $\Gamma$ to close the artificial bandgap between the first two dispersion curves created by the shift by \(\Delta\).
    \item \emph{Estimation}: The corrected spectrum is then used as input for the trained DeepF-fNet to compute the optimal parameters.
\end{enumerate}

By adhering to the steps described herein, the trained DeepF-fNet model can be utilized effectively to determine, in real time, the optimal parameters of metamaterials designed to exclude a specified frequency. Algorithm \ref{alg:SICE} presents the most comprehensive iteration of the SICE4 algorithm, in which the target frequencies are introduced as a time series, thus mirroring the conditions commonly observed in practical applications.
\begin{algorithm}[h!]
	\caption{SICE4 algorithm}
	\label{alg:SICE}
	\begin{algorithmic}[1]
		\STATE{$\underline{f}_{bg}=average\,1^{st}\,bandgap\,frequencies$}
		\STATE{$\hat{\underline{f}} = desired\,frequencies\,to\,stop\,as\,a\,vector$}
		\FOR{$t\leq len\left(\hat{\underline{f}}\right)$}
		\STATE{$\underline{MSE}=\dfrac{1}{len\left(\underline{f}_{bg}\right)}\,\sum_{i=1}^{len\left(\underline{f}_{bg}\right)}\left(\underline{f}_{bg}-\hat{\underline{f}}\left[t\right]\right)^{2}$}
		\STATE{$ib=argmin\left(\underline{MSE}\right)$}
		\STATE{$\Delta=\hat{\underline{f}}\left[t\right]-\underline{f}_{bg}\left[ib\right]$}
		\STATE{$new\_DC_{n>1}=DC\left[ib\right]+\Delta$}
		\STATE{$\Gamma=min\left(new\_DC_{n=2}\right)/ max\left(DC_{n=1}\left[ib\right]\right)$}
		\STATE{$new\_DC_{n=1}=\Gamma \cdot DC_{n=1}\left[ib\right]$}
		\STATE{$Parameters=IEPS\left(new\_DC\right)$}
		\ENDFOR
	\end{algorithmic}
\end{algorithm}

\subsection{DeepF-fNet Framework}
\label{subsec:DeepF-fNet}
\begin{figure}[h!]
    \centering
    \includegraphics[width=1\linewidth]{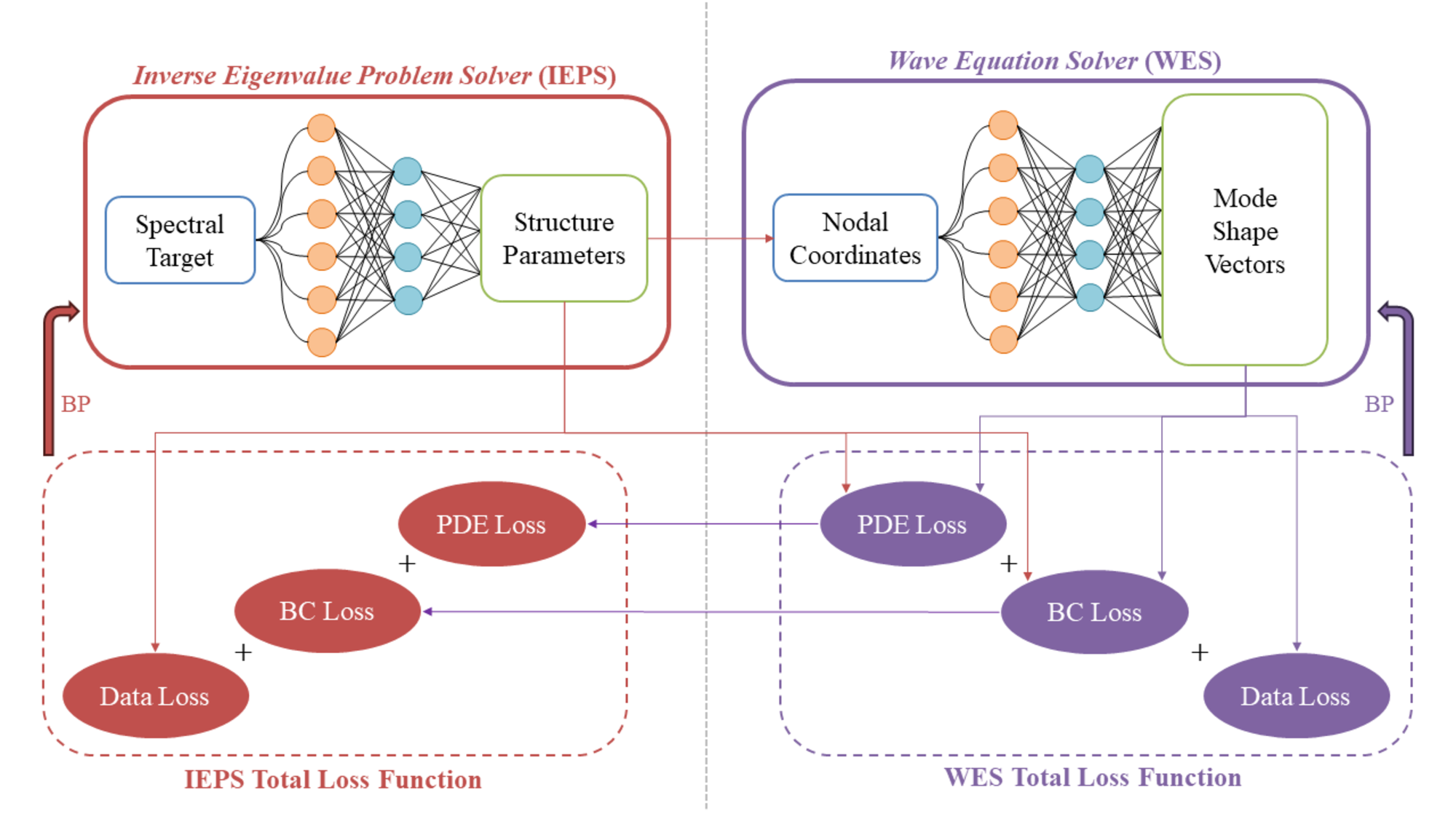}
    \caption{DeepF-fNet architecture}
    \label{fig:deepF-fnetarchitecture}
\end{figure}

DeepF-fNet is structured through the dual network configuration shown in Fig.~\ref{fig:deepF-fnetarchitecture}: the inverse eigenvalue problem solver (IEPS) and the wave equation solver (WES). The IEPS is responsible for estimating structural parameters aligned with a desired frequency spectrum, while the WES is tasked with computing eigenvectors pertinent to a metamaterial configuration to ensure adherence to physical accuracy. The framework is named after its bifurcated architecture: the IEPS models a functional by mapping scalars from the function set $\omega_{n}(\kappa)$, whereas the WES approximates a function, specifically targeting the eigenfunction of the given structure.

Both networks of DeepF-fNet are convolutional neural networks (CNNs). This architecture was chosen for its effectiveness in learning from 2D data (e.g., matrices). The IEPS takes in the set of $N$ dispersion curves and outputs the set of structural parameters of the unit cell whose response is the target spectrum. The WES processes the unit cell generated by the IEPS, which is discretized into sampling points. At these points, the reconstruction loss is minimized while simultaneously enforcing the physical laws governing the eigenfrequency problem. Therefore, the WES must be located after the IEPS in the processing sequence. The WES output is a set of $\Tilde{N}$ mode shapes, defined as the displacement vectors at the discretization points.

The enforcement of the problem's physics is achieved by integrating the outputs of both models in a physically coherent manner throughout the training phase. The comprehensive loss function is composed of three distinct components: a data-driven element, termed the data loss, and two physics-informed elements, encompassing the partial differential equation (PDE) loss and the boundary condition (BC) loss. Together, these elements contribute to the total loss function of the $j^{th}$ network, as depicted in Eq.~\ref{eq:totalloss}.
\begin{equation}
	\label{eq:totalloss}
		\mathcal{L}_{j}^{\ast}=\dfrac{1}{N_{j,o}}\,\sum_{i=1}^{N_{j,o}}\left(\hat{y}_{j,i}-y_{j,i}\right)^{2}
		+\dfrac{1}{N_{PDE}}\,\sum_{i=1}^{N_{PDE}}f_{PDE}^{2}+\dfrac{1}{N_{BC}}\,\sum_{i=1}^{N_{BC}}f_{BC}^{2}
\end{equation}
The first term in Eq.~\ref{eq:totalloss} quantifies the reconstruction error between the predicted output $y_{j,i}$ and the ground truth $\hat{y}_{j,i}$. This measure is crucial to assess the accuracy of the prediction compared to the actual data. The second term refers to the residual $f_{PDE}$ associated with the governing harmonic wave equation, as referenced in Eq.~\ref{eq:governing_equation} \cite{comsol}. This term is integral to enhancing the learning process by incorporating both the material density $\rho$ and the elasticity tensor $\underline{\underline{C}}$, which are functions of the structural parameters, along with the eigenvectors $\underline{w}$. Therefore, the PDE loss is necessarily computed on the basis of the prediction of the WES. Lastly, the third component enforces the boundary conditions (BCs) of the PDE through the residual $f_{BC}$.
\begin{equation}
	\label{eq:governing_equation}
	4\,\pi^{2}\,\rho \, f_{n}^{2}\, \underline{w}+\nabla \cdot \left[\underline{\underline{C}}:\dfrac{1}{2}\left(\nabla \underline{w}^{T}+\nabla \underline{w}\right)\right]=\underline{0}
\end{equation}
Although data losses pertain to each individual network, the PDE and BC losses are common to both the IEPS and WES. Consequently, even though these losses are formulated after the second network, they concurrently influence the training dynamics of the first network.

During the training phase, the weights of the CNNs are optimized using the \emph{Adam} optimizer \cite{kingma2017adammethodstochasticoptimization}. This optimizer is responsible for computing and applying the gradients of the total loss function with respect to the network weights. To further enhance the training process, a dynamic reduction in the learning rate is implemented through the ReduceLROnPlateau callback \cite{tensorflow2015-whitepaper} as the training progresses.

Because the gradients in both networks have different magnitudes, batch normalization \cite{ioffe2015batchnormalizationacceleratingdeep} is utilized in conjunction with norm clipping \cite{Pascanu2012UnderstandingTE} and dropout \cite{tensorflow2015-whitepaper} to prevent gradient explosion and vanishing.

%% file: sections/casestudy.tex
\section{Case Study}
\label{sec:case study}
The proposed framework was validated through a case study inspired by the work of Jung et al. \cite{LRMHyundai}, involving a locally resonant metamaterial (LRM) engineered to attenuate vibrations on a steel plate. This case was selected to rigorously evaluate the effectiveness of the framework in addressing complex vibrational challenges and to demonstrate its applicability to optimize metamaterial configurations to damp out unwanted frequencies.
\subsection{Problem Definition}
\label{subsec:problem definition}
\begin{figure}[h!]
    \centering
    \begin{subfigure}{0.49\textwidth}
    \includegraphics[width=0.95\linewidth]{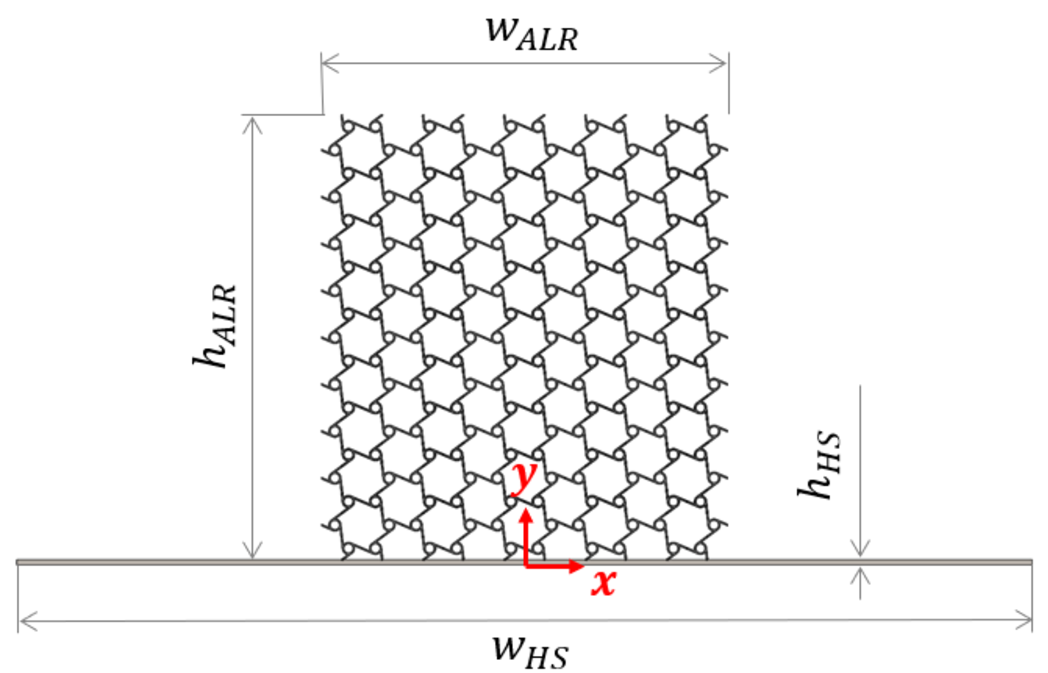}
    \caption{}
    \label{fig:rendering}
    \end{subfigure}
    \begin{subfigure}{0.49\textwidth}
    \includegraphics[width=0.5\linewidth]{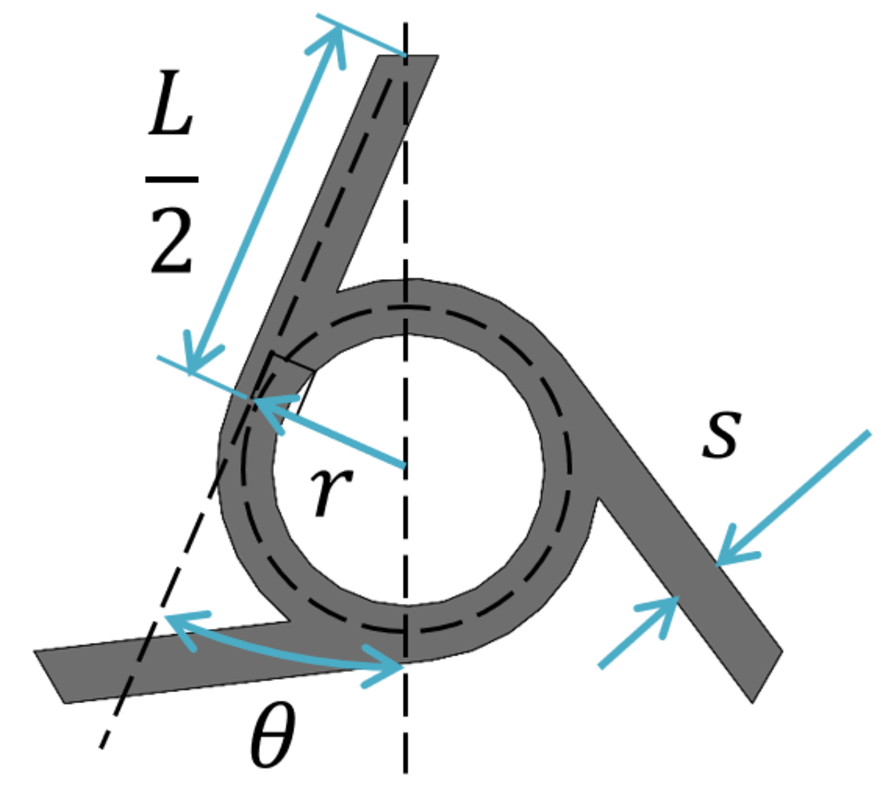}
    \caption{}
    \label{fig:unit-cell}
    \end{subfigure}\\
    \begin{subfigure}{1\textwidth}
    \includegraphics[width=1\linewidth]{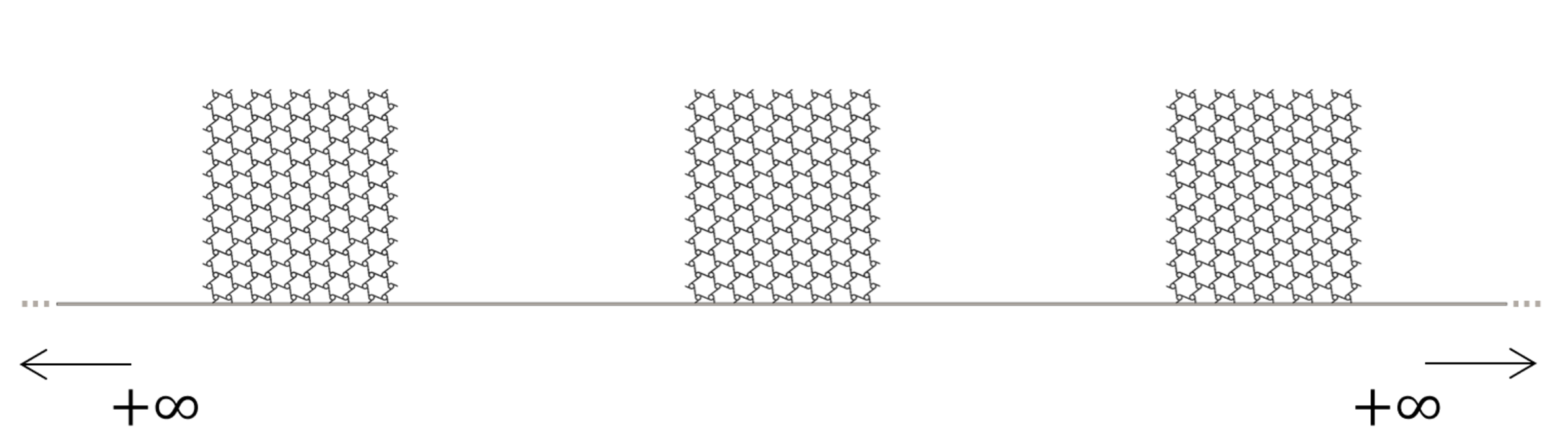}
    \caption{}
    \label{fig:periodic}
    \end{subfigure}
    \caption{LRM geometry: (a) unit cell, (b) ALR's unit cell and (c) periodic arrangement}
    \label{fig:LRM}
\end{figure}

The unit cell of the examined LRM consists of two main components: a host structure (HS), which is the primary target for vibration suppression, and an attachable local resonator (ALR), which serves as the semi-active medium for vibration isolation (Fig.~\ref{fig:rendering}) \cite{LRMHyundai}.

The ALR itself is an LRM, with its design featuring a trichiral honeycomb structure selected for its proven effectiveness in facilitating bandgap formation \cite{YU2018114,ABDELJABER201650,BARAVELLI20136562,MOUSANEZHAD201681}. The isotropic pattern is defined by three independent parameters — $r$, $L$ and $s$ — and the angle $\theta=atan\left(\frac{2r}{L}\right)$ (Fig.~\ref{fig:unit-cell}). The proposed pattern is two-dimensional.  Subsequently, under the assumption of a plane-strain condition, it is extruded along the axis perpendicular to its plane to achieve a three-dimensional form. In order to have a robust geometry of the ALR pattern, the three independent parameters are subjected to compliance constraints (Eq.~\ref{compliance}). These conditions are imposed during data generation and are monitored during training, to prevent the formation of self-intersecting or degenerate geometries.
\begin{subequations}
	\label{compliance}
	\begin{align}[left=\empheqlbrace]
		r&>\dfrac{s}{2} \label{compliance1}\\
		\dfrac{L}{2}&>r+\dfrac{s}{2} \label{compliance2}
	\end{align}
\end{subequations}

The height of the HS (along the y-axis shown in Fig.~\ref{fig:rendering}) was fixed at \(0.8 \, \text{mm}\), while its width (along the y-axis) was optimized based on the width ratio between the resonator and the underlying plate, crucial for effective vibration isolation. Proper spacing creates destructive interference for certain frequency ranges of interest, improving vibration reduction without excessively adding mass, as noted by Claeys et al. \cite{CLAEYS20131418}.

The LRM unit cell is periodically extended in both the left and right directions, as shown in Fig.~\ref{fig:periodic}.
The geometric boundaries of the LRM unit cell are defined by four dimensions, expressed in Eq.~\ref{size}.
 \begin{subequations}
 	\label{size}
 	\begin{align}
		w_{ALR}=n_{H}\cdot\dfrac{3\,L}{cos(\theta)}\label{size1}\\
		h_{ALR}=n_{V}\cdot\dfrac{\sqrt{3}\,L}{cos(\theta)}\label{size2}\\
		w_{HS}=2.5\,w_{ALR}\\
		h_{HS}=0.8\,mm
 	\end{align}
 \end{subequations}
In this context, $w$ and $h$ indicate width (along the x-axis shown in Fig.~\ref{fig:rendering}) and height (along the y-axis shown in Fig.~\ref{fig:rendering}), respectively. To derive Eqs.~\ref{size1} and~\ref{size2}, the unit cells of the ALR have been grouped into regular hexagons, where the parameters $n_{H}$ and $n_{V}$ denote the number of hexagonal clusters repeated along the horizontal and vertical directions, respectively. In this study, $n_{H}=5$ and $n_{V}=9$, creating resonators with balanced dimensions across both axes.

 The mechanical properties of the ALR depend on the geometric parameters shown in Eq.~\ref{Epv} \cite{MOUSANEZHAD201681}.
 \begin{subequations}
	\label{Epv}
	\begin{align}
		\frac{\rho_{ALR}}{\rho} &= \frac{2}{\sqrt{3}} \left(\frac{s}{L}\right) \left[\frac{1 + \frac{4\pi}{3} \cdot \left(\frac{r}{L}\right)}{1 + 4\left(\frac{r}{L}\right)^2}\right]\\
		\frac{E_{ALR}}{E} &= \frac{4}{\sqrt{3}} \left(\frac{s}{L}\right)^3 \left[\frac{\frac{3}{2}}{\cos^2\left(\frac{\pi}{6} - \theta\right) + 4 \sin^2(\theta) + \cos^2\left(\frac{\pi}{6} + \theta\right)}\right]\\
		\nu_{ALR} &= \sqrt{3} \left[\frac{\sin\left(\frac{\pi}{6} - \theta\right)\cos\left(\frac{\pi}{6} - \theta\right) + \sin\left(\frac{\pi}{6} + \theta\right)\cos\left(\frac{\pi}{6} + \theta\right)}{\cos^2\left(\frac{\pi}{6} - \theta\right) + 4 \sin^2(\theta) + \cos^2\left(\frac{\pi}{6} + \theta\right)}\right]
	\end{align}
\end{subequations}
In this context, $\rho_{ALR}$, $E_{ALR}$ and $\nu_{ALR}$ are the effective density, elastic modulus, and Poisson's ratio of the ALR, respectively, while $\rho$ and $E$ are the corresponding properties of the bulk material. The bulk material properties of ALR and HS are reported in Tab.~\ref{table:mat_prop}.
 \begin{table}[H]
 	\centering 
 	\begin{tabular}{c c c c}
 		\hline
 		& $\rho\,\left[kg/m^{3}\right]$ & $E\,\left[GPa\right]$ & $\nu\,\left[-\right]$\\
 		\hline \hline
 		\textbf{ALR}: PA12 nylon & 1010 & 1.215 & -\\
 		\textbf{HS}: C30 steel & 7850 & 212 & 0.29\\
 		\hline
 	\end{tabular}
 	\\[10pt]
 	\caption[LRM material properties]{Relevant bulk material properties of the LRM}
 	\label{table:mat_prop}
 \end{table}
The robustness of the LRM model is grounded in the validity of the Floquet-Bloch theorem (Eq.~\ref{FloqBloch} \cite{COLLET20112837}), where $\underline{w}_{0}$ is the amplitude of a generic harmonic wave, while $\underline{w}_{n,\kappa}$ is the $\Omega_{R}$-periodic, space-dependent displacement function. This function depends on the considered $n^{th}$ harmonic component and the $\kappa^{th}$ wave number. 
\begin{subequations}
\label{FloqBloch}
\begin{align}[left=\empheqlbrace]
\underline{w}_{0}\left(x\right)=\underline{w}_{n,\kappa}\left(x,\kappa\right)e^{-i\,\kappa\,x}\,\,\,\forall x\in \Omega_{R}\\
\underline{w}_{n,\kappa}\left(x-d\right)=\underline{w}_{n,\kappa}\left(x\right)\,\,\,\forall x \in \partial \Omega_{R} 
\end{align}
\end{subequations}
Furthermore, \(\Omega_{R}\) represents the periodically repeating domain of length \(d\) along the \(x\) axis (in this case study, $d=w_{HS}$). This domain reflects the uniquely defined primitive cell in the reciprocal space, also known as the first Brillouin zone, whose analysis suffices to characterize the wave propagation properties throughout the periodic medium \cite{COLLET20112837}.

Three boundary conditions are applied:
 \begin{itemize}
     \item \emph{Floquet periodicity} on the left and right edges of the HS. This condition enforces the periodicity of the LRM unit cell according to Eq.~\ref{FlochPer1D}  \cite{comsol}.
    \begin{equation}
 	\label{FlochPer1D}
 	\underline{w}_{right}=\underline{w}_{left}\cdot e^{-i\,\kappa \,w_{HS}}
    \end{equation}
     \item \emph{Displacement and force continuity} at the interface between the HS and the ALR.
     \item \emph{Traction-free surfaces} on the remaining boundaries.
 \end{itemize}

As already mentioned in \autoref{sec:methodology}, the WES model takes in a discretized geometry to compute the eigenvectors. In this case study, the LRM was divided into 51 sampling nodes defined by the parametric grid represented in Tab~\ref{tab:coord}. Here, the top row refers to the coordinate $x$, while the left column refers to the coordinate $y$. The number inside each cell represents the label of the corresponding node. Blank cells mean that no node is present at that location.
\begin{table}[h!]
\centering
\begin{tabular}{c|>{\centering\arraybackslash} p{3em}>{\centering\arraybackslash} p{3em}>{\centering\arraybackslash} p{3em}>{\centering\arraybackslash} p{3em}>{\centering\arraybackslash} p{3em}>{\centering\arraybackslash} p{3em}>{\centering\arraybackslash} p{3em}}
  		&$-\frac{w_{HS}}{2}$ & $-\frac{w_{ALR}}{2}$ &$ -\frac{w_{ALR}}{4}$ & $0$ & $\frac{w_{ALR}}{4}$ & $\frac{w_{ALR}}{2}$ & $\frac{w_{HS}}{2}$\\
  		\hline
  		$h_{HS}+h_{ALR}$&&47&48&49&50&51&\\
  		$h_{HS}+\frac{5\,h_{ALR}}{6}$&&42&43&44&45&46&\\
  		$h_{HS}+\frac{2\,h_{ALR}}{3}$&&37&38&39&40&41&\\
  		$h_{HS}+\frac{h_{ALR}}{2}$&&32&33&34&35&36&\\
  		$h_{HS}+\frac{h_{ALR}}{3}$&&27&28&29&30&31&\\
  		$h_{HS}+\frac{h_{ALR}}{6}$&&22&23&24&25&26&\\
  		$h_{HS}$&15&16&17&18&19&20&21\\
  		$\frac{h_{HS}}{2}$&8&9&10&11&12&13&14\\
  		$0$&1&2&3&4&5&6&7\\
  		\hline
\end{tabular}
\\[10pt]
\caption{List of the parametric nodal coordinates. the top row refers to the coordinate $x$, while the left column refers to the coordinate $y$. The number inside each cell represents the label of the corresponding node. Blank cells mean that no node is present at that location.}
\label{tab:coord}
\end{table}

The DeepF-fNet parameters defining the architecture, the optimizer setup, and the weights of the loss functions for this case study are summarized in Tabs.~\ref{paramNet}, ~\ref{paramOpt}, and~\ref{weights}.
 \begin{table}[h!]
 	\centering 
 	\begin{tabular}{c  c  c }
 		\hline
 		& \textbf{IEPS} & \textbf{WES} \\
 		\hline
 		\textbf{Input size} & $10\times 40$ & $81600\times 2$\\
 		\textbf{Output size} & $3\times 1$ & $81600\times 1$\\
 		\textbf{Convolutional layers} & $3$ & $2$\\
 		\textbf{Convolutional filter size} & $3\times 3$ &$2\times 2$ \\
 		\textbf{Stride} & $1$ &$1$ \\
 		\textbf{Padding} & \emph{same} & \emph{same} \\
 		\textbf{Max pooling size} & $2\times 2$ &$-$ \\
 		\textbf{Fully connected layer size} & $160\times 1$ &$20\times 1$ \\
		\hline
 	\end{tabular}
 	\\[10pt]
 	\caption{DeepF-fNet architecture}
 	\label{paramNet}
 \end{table}
 \begin{table}[h!]
	\centering 
	\begin{tabular}{c  c  c }
		\hline
		& \textbf{IEPS} & \textbf{WES} \\
		\hline
 		\textbf{Initial learning rate} & $10^{-3}$ & $10^{-2}$\\
		\textbf{Minimum delta} & $10^{-10}$ & $10^{-16}$\\
		\textbf{Reduction factor} & $0.5$ & $0.5$\\
		\textbf{Patience} & $1$ & $1$\\
		\hline
	\end{tabular}
	\\[10pt]
	\caption{\emph{Reduce on plateau} settings for \emph{Adam} optimizer}
	\label{paramOpt}
\end{table}
\begin{table}[h!]
\centering 
\begin{tabular}{c  c  c }
	\hline
	& \textbf{IEPS} & \textbf{WES} \\
	\hline
	\textbf{Data Loss} & $10^{8}$ & $10^{12}$\\
	\textbf{PDE Loss} & $10^{-14}$ & $10^{-14}$\\
	\textbf{BC Loss} & $10^{-12}$ & $10^{-12}$\\
	\hline
\end{tabular}
\\[10pt]
\caption{Weights of the loss functions}
\label{weights}
\end{table}
Each row of the IEPS input corresponds to a specific eigenfrequency of the dispersion curves, while each column corresponds to a division of the wave number range $\left[-\frac{\pi}{w_{HS}}, \frac{\pi}{w_{HS}}\right)$. Thus, the i-th row specifies the i-th eigenfrequency, and the j-th column specifies the j-th division within the given range. The size of the WES input and of its output is chosen to effectively apply the PDE residual. In the two-dimensional domain, each of the 51 nodes has two displacement components: $u$ and $v$. These components must be calculated for all $n$ and $\kappa$, resulting in 400 different scenarios (computed as $10 \; n \cdot 40 \; \kappa$). Due to the Floquet-Bloch theorem, both real and imaginary parts of these displacement components must be considered. This results in a total of $81600$ nodal displacements, which constitute the WES output. The calculation of these displacements is derived as follows: $51$ nodes, each with $2$ displacement components (for $u$ and $v$), are evaluated in $10$ different scenarios (variable $n$) and $40$ wave numbers (variable $\kappa$), with both the real part and imaginary part considered, hence multiplication by $2$. Consequently, the input is made up of coordinate pairs $(x, y)$ corresponding to the nodes, structured as a vector of dimensions $81600 \times 2$.

Regarding the remaining network parameters, a dropout rate of 0.5 is applied to prevent overfitting, while norm-clipping is enforced with a threshold of 1 to stabilize gradient updates. The activation function used in the hidden layers is the hyperbolic tangent (\emph{tanh}), with a linear activation function in the output layer to support continuous value predictions. Standardization incorporates a bias term \(\epsilon = 10^{-7}\) for numerical stability, according to the batch normalization technique proposed in Ref.~\cite{ioffe2015batchnormalizationacceleratingdeep} . The dataset comprises 2000 samples, divided into 1000 for training, 501 for validation, and 499 for testing. A mini-batch size of 40 is employed over 40 epochs to optimize network learning.
\subsection{Dataset Generation}
\label{subsec:dataset generation}
The dataset used to train, validate, and test the framework was generated through FE simulations. MATLAB\textsuperscript{\textregistered} served as a scripting interface to set up and solve models in COMSOL\textsuperscript{\textregistered}, based on a set of randomly selected independent structural parameters (that is, $r$, $L$ and $s$) between $0.04\,mm$ and $4\,mm$. The two packages communicate with each other through the LiveLink\textsuperscript{\textregistered} server.

The structure was modeled as a junction of two rectangular domains (that is, ALR and HS, according to Eq.~\ref{size}) with the respective material properties defined by Eq.~\ref{Epv} and Tab.~\ref{table:mat_prop}.

The domain was discretized using a physics-controlled mesh of size \emph{Finer} (that is, the third most refined mesh out of nine levels \cite{comsol}). The element count for this mesh type varies according to the absolute size of the generated geometry, while maintaining a nearly constant refinement level. This mesh size was selected on the basis of a convergence study, providing an optimal balance between computational cost and solution accuracy. Triangular elements with second-order integration were used \cite{comsol}.

The mode shapes for any discrete wave number $K\in \left[-\frac{\pi}{w_{HS}},\frac{\pi}{w_{HS}}\right)$ were determined by a linear eigenfrequency study, where the unforced undamped structure is governed by Eq.~\ref{eq:governing_equation} \cite{comsol}. The adopted default solver was ARPACK, with a manual eigenfrequency search method, which features the following settings:
\begin{itemize}
    \item \emph{Desired number of eigenfrequencies} (i.e., $N$): 10
    \item \emph{Search for eigenvalues around}: $1\,Hz$ (default)
    \item \emph{Eigenfrequency search method around shift}: closest in absolute value
\end{itemize}
The boundary conditions described in \autoref{subsec:problem definition} were implemented. In addition, combinations of parameters that do not comply with the constraints described in Eq.~\ref{compliance} were discarded.

The pseudo-code describing dataset generation is reported in Algorithm \ref{alg:datagen}.
\begin{algorithm}[h!]
	\caption{Data generation}
	\label{alg:datagen}
	\begin{algorithmic}[1]
		\STATE{Define the dataset $size$}
		\STATE{Define a $range$ for the structural parameters}
		\FOR{$j\leq size$}
		\STATE{$parameters_{j}=random\left(range\right)$}
		\IF{$constraints\left(parameters_{j}\right)==True$}
		\STATE{save $parameters_{j}$ $\rightarrow$ IEPS labels}
		\STATE{construct sampling grid using $parameters_{j}$}
		\STATE{run Comsol\textsuperscript{\textregistered} Multi-physics\textsuperscript{\textregistered} study with $parameters_{j}$}
		\STATE{save eigenfrequencies $\forall n$ and $\forall \kappa$ $\rightarrow$ dispersion curve}
		\STATE{save displacement field $@$ sampling grid $\forall n$ and $\forall \kappa$ $\rightarrow$ WES labels}
		\ENDIF
		\ENDFOR
		\STATE{export the constructed tensor as $Dataset$}
	\end{algorithmic}
\end{algorithm}

The dataset encompasses the following data pertinent to each simulation:
\begin{itemize}
    \item \emph{Structural Parameter Labels}: These parameters define the numerical simulations and are compared to the IEPS predictions to calculate the data-driven loss component.
    \item \emph{Ground-truth Dispersion Curves}: Arranged in an $N\times K$ matrix, these numerical dispersion curves are utilized within the IEPS framework.
    \item \emph{Mode Shape Labels}: Represented as $N\cdot K$ eigenvectors $\underline{w}$, these mode shapes are compared with WES predictions to enforce the physics-informed loss.
\end{itemize}

\subsection{Results}
\label{subsec:results}
\begin{figure}[h!]
	\centering
	\subfloat[IEPS\label{fig:loss_IEPS}]{
		\includegraphics[scale=0.49]{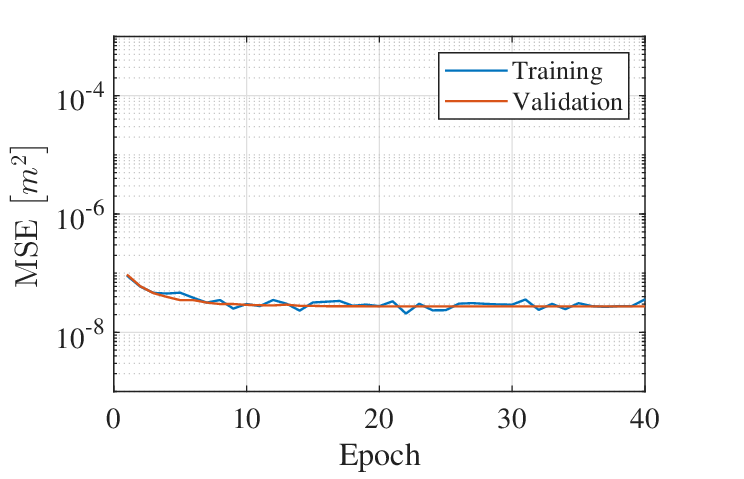}
	}
	\quad
	\subfloat[WES\label{fig:loss_WES}]{
		\includegraphics[scale=0.49]{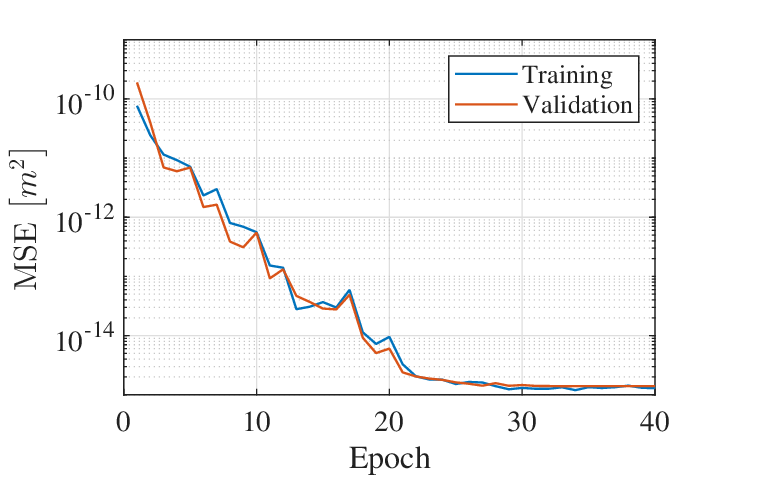}
	}
	\caption[Training and validation losses]{Training and validation losses of the IEPS and WES, computed with respect to the labeled data}
	\label{fig:loss}
\end{figure}

\begin{figure}[h!]
	\quad
	\subfloat[IEPS total loss\label{fig:total_IEPS}]{
	\includegraphics[scale=0.47]{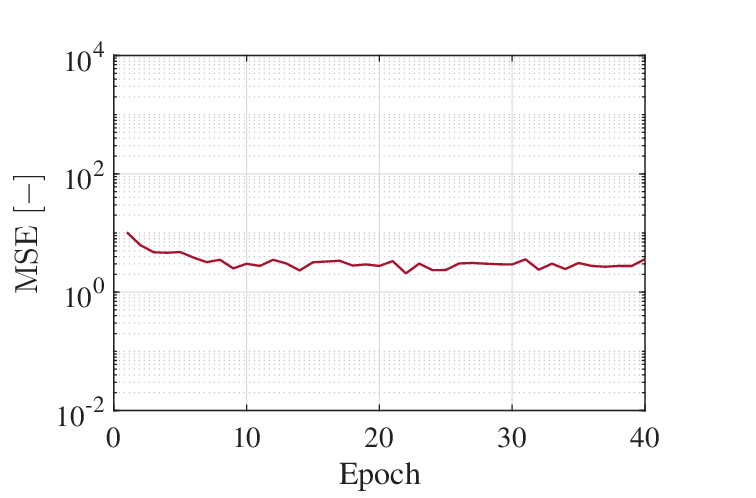}
	}
	\quad
	\subfloat[WES total loss\label{fig:total_WES}]{
	\includegraphics[scale=0.47]{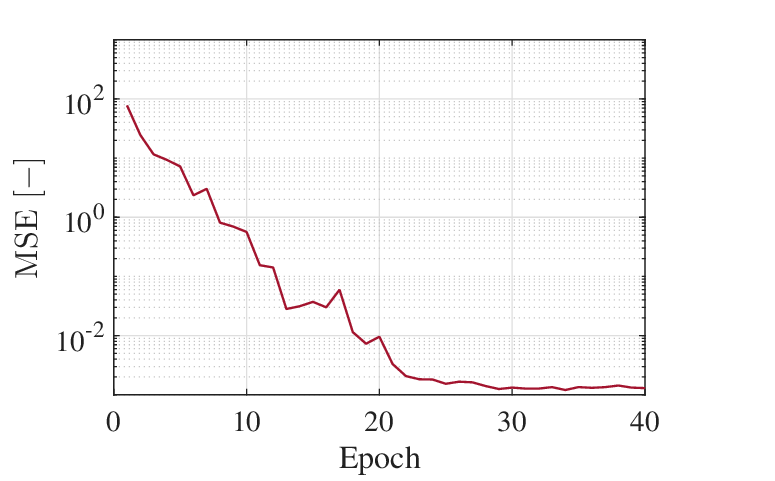}
	}\\
	\centering
	\subfloat[PDE loss\label{fig:PDE_loss}]{
	\includegraphics[scale=0.47]{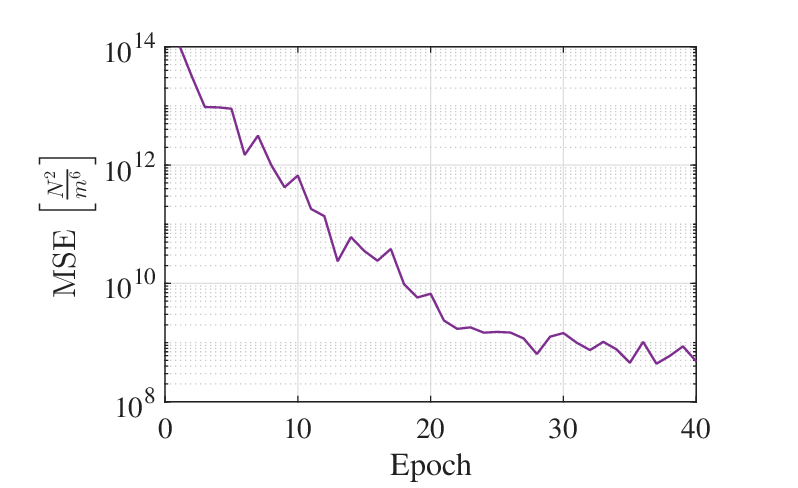}
	}
	\quad
	\subfloat[BC loss\label{fig:BC_loss}]{
	\includegraphics[scale=0.47]{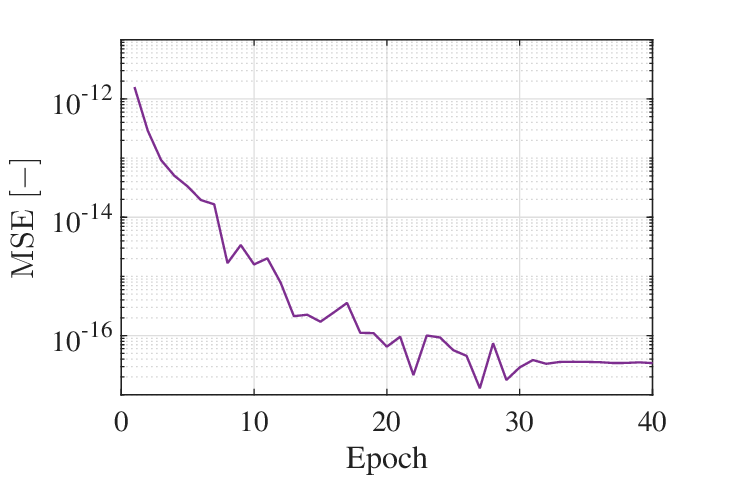}
	}\\
	\quad
	\subfloat[Compliance conditions\label{fig:manu_loss}]{
	\includegraphics[scale=0.47]{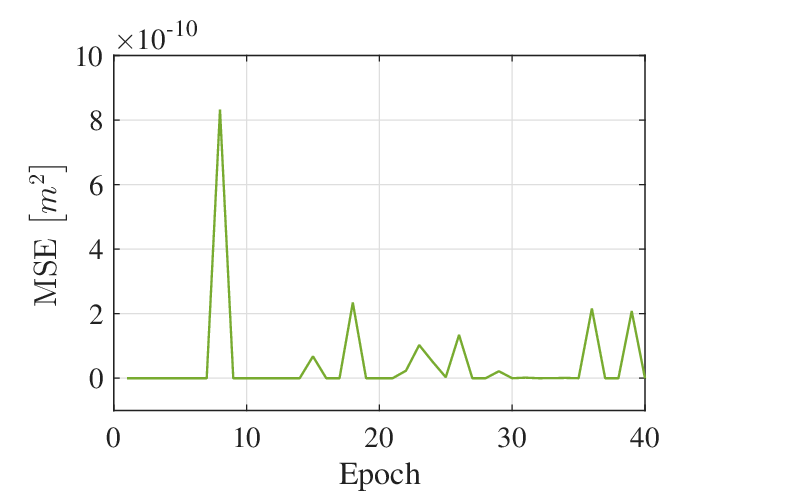}
	}
	\caption[Loss components]{Total IEPS and WES loss functions with their physics-informed components, calculated on the training dataset}
	\label{fig:losses}
\end{figure}

The DeepF-fNet framework training process is shown in Figs. \ref{fig:loss} and \ref{fig:losses}. In Fig. \ref{fig:loss}, the data loss is compared between the training and validation sets, while Fig. \ref{fig:losses} presents all components of the total loss functions, including the total losses.
All loss functions and associated components decrease, and the IEPS validation loss converges much sooner than the corresponding WES. Although showing a small drop in the plot, the IEPS shows a final MSE with respect to the labeled data of $2.74\cdot 10^{-8}\,m^{2}$, corresponding to a root mean square deviation of the estimated parameters of $0.166\,mm$. Regarding WES, the MSE of the labeled data amounts to $1.40\cdot 10^{-15}\,m^{2}$, equivalent to $0.0374\,\mu m$ of the root mean square deviation of the nodal displacements. Despite being a very small deviation, it is common for this LRM to manifest maximum displacements of the order of $10^{-1}\div 10^{-4}\,\mu m$, indicating a quite scattered behavior of the error. The compliance conditions (Fig.~\ref{fig:manu_loss}) are violated for some combinations of parameters, but it was empirically observed that these outliers do not compromise the consistency of the predictions of the trained network.

\begin{figure}[h!]
	\centering
	\subfloat[\label{fig:IEPSval1}]{
		\includegraphics[scale=0.47]{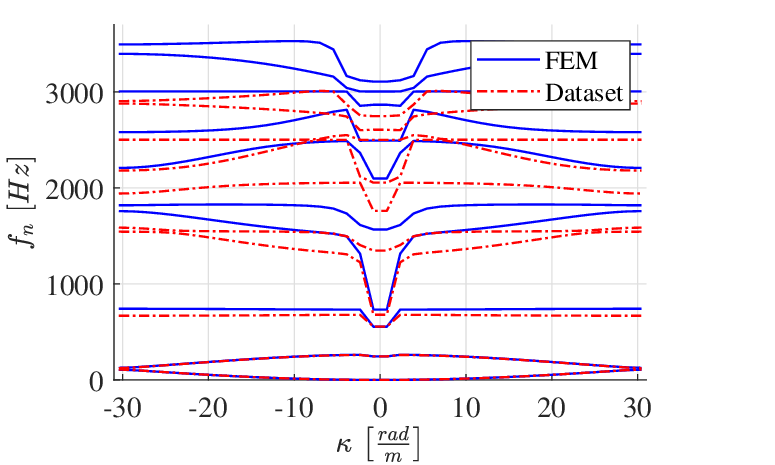}
	}
	\quad
	\subfloat[\label{fig:IEPSval12}]{
		\includegraphics[scale=0.47]{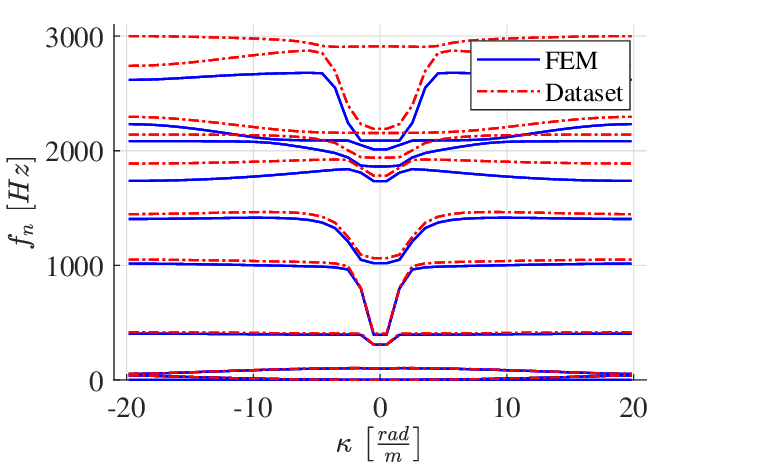}
	}\\
	\subfloat[\label{fig:IEPSval2}]{
		\includegraphics[scale=0.47]{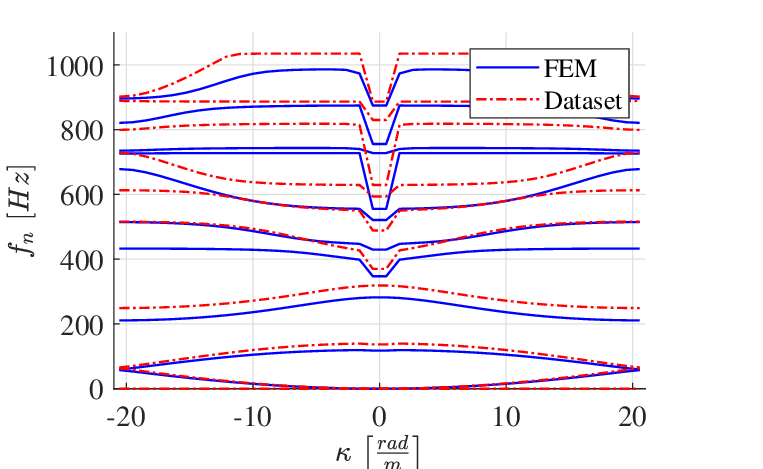}
	}
	\quad
	\subfloat[\label{fig:IEPSval22}]{
		\includegraphics[scale=0.47]{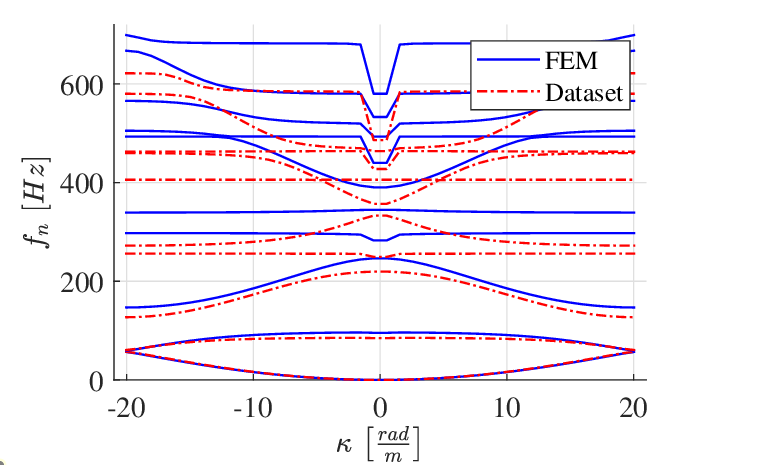}
	}\\
	\subfloat[\label{fig:IEPSval3}]{
		\includegraphics[scale=0.47]{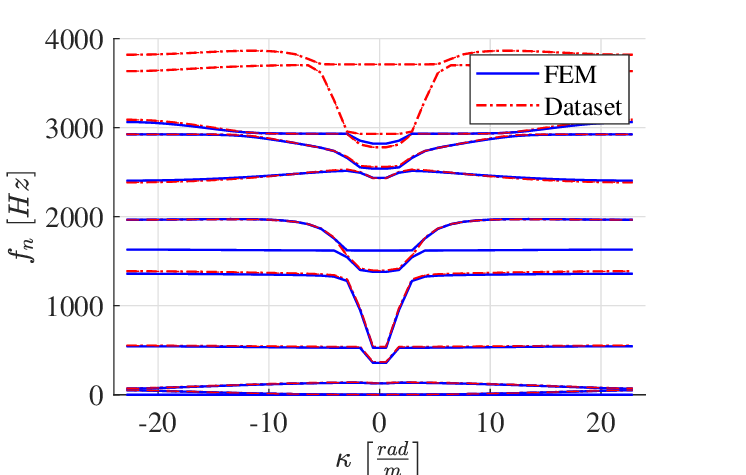}
	}
	\quad
	\subfloat[\label{fig:IEPSval32}]{
		\includegraphics[scale=0.47]{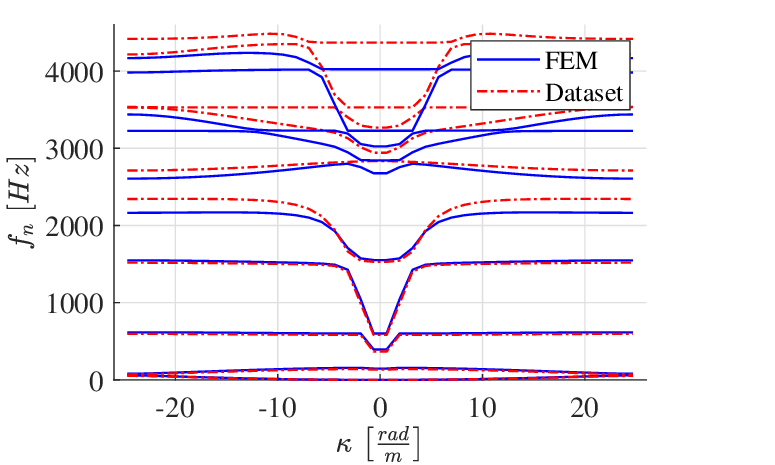}
	}
	\caption[Validation IEPS model]{Graphic validation of the IEPS model with input dispersion curves coming from the  validation (a,b), test (c,d), and training (e,f) splits}
	\label{fig:IEPSval}
\end{figure}

To validate the IEPS it is crucial to plot the dispersion curves associated with the predicted parameters versus the actual ones that the network took in. For this purpose, a sample was collected from each of the three splits and fed to the trained model. The predicted parameters were then prompted to the FE solver of the direct problem to obtain the associated dispersion curves. The comparison between the two sets of curves can be observed in Fig. \ref{fig:IEPSval} and Tab. \ref{tab:disp}.
 \begin{table}[h!]
	\centering 
	\begin{tabular}{c c  c  c c}
		\hline
		& & \textbf{True} & \textbf{Predicted} & \textbf{Error} \\
		\hline \hline
		& \textbf{r} $\bm{\left[mm\right]}$ & $0.800$ & $0.762$ & $4.75\,\%$\\
		(a) & \textbf{L} $\bm{\left[mm\right]}$ & $2.28$ & $2.29$&$0.439\,\%$\\
		& \textbf{s} $\bm{\left[mm\right]}$ & $0.560$ & $0.658$&$17.5\,\%$\\
		\hline
			& \textbf{r} $\bm{\left[mm\right]}$ & $0.800$ & $0.910$&$13.8\,\%$\\
			(b) &\textbf{L} $\bm{\left[mm\right]}$ & $3.92$ & $3.81$&$2.81\,\%$\\
			& \textbf{s} $\bm{\left[mm\right]}$ & $1.36$ & $1.29$&$5.15\,\%$\\
			\hline
			& \textbf{r} $\bm{\left[mm\right]}$ & $0.800$ & $1.02$&$27.5\,\%$\\
			(c) & \textbf{L} $\bm{\left[mm\right]}$ & $3.60$ & $3.53$&$1.94\,\%$\\
			& \textbf{s} $\bm{\left[mm\right]}$ & $0.360$ & $0.342$&$5.00\,\%$\\
			\hline
			& \textbf{r} $\bm{\left[mm\right]}$ & $1.24$ & $1.09$&$12.1\,\%$\\
			(d) &\textbf{L} $\bm{\left[mm\right]}$ & $3.32$ & $3.55$&$6.93\,\%$\\
			& \textbf{s} $\bm{\left[mm\right]}$ & $0.200$ & $0.245$&$22.5\,\%$\\
			\hline
			& \textbf{r} $\bm{\left[mm\right]}$ & $0.800$ & $0.785$&$1.88\,\%$\\
			(e) &\textbf{L} $\bm{\left[mm\right]}$ & $3.24$ & $3.31$&$2.16\,\%$\\
			& \textbf{s} $\bm{\left[mm\right]}$ & $1.32$ & $1.37$&$3.79\,\%$\\
			\hline
			& \textbf{r} $\bm{\left[mm\right]}$ & $0.800$ & $0.775$&$3.13\,\%$\\
			(f) & \textbf{L} $\bm{\left[mm\right]}$ & $3.12$ & $3.02$&$3.21\,\%$\\
			& \textbf{s} $\bm{\left[mm\right]}$ & $1.48$ & $1.27$&$14.2\,\%$\\
			\hline
		\end{tabular}
	\caption[IEPS numerical validation]{Numerical comparison between the IEPS ground truth and predicted structural parameters. Each table is referenced to the corresponding dispersion curves in Fig. \ref{fig:IEPSval}}
	\label{tab:disp}
\end{table}
Although not all curves are well approximated by the IEPS model, a common feature is the good accuracy of the position of the first bandgap, typically between the 2\textsuperscript{nd} and the 3\textsuperscript{rd} eigenfrequency. It can also be noted that the prediction of the higher eigenfrequencies shows a prominent error with respect to the input dispersion curves: the spectral bias of the high-frequency components is not uncommon in PINNs for inverse problems \cite{HE2024117130,HU2024127240}, but since the widest bandgap has been empirically observed to be at the lowest frequencies, it does not affect the reliability of SICE4.

\begin{figure}[h!]
	\centering
	\subfloat[Ground truth\label{fig:WESval1}]{
		\includegraphics[scale=0.48]{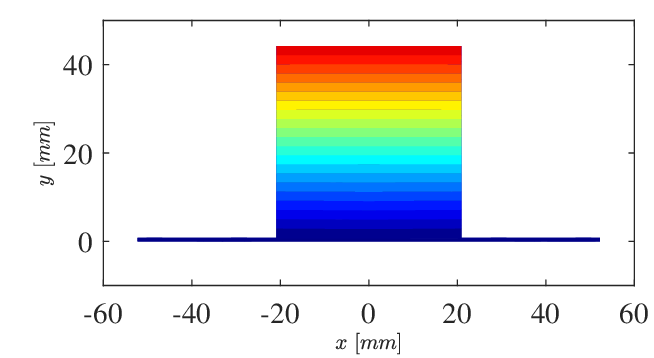}
	}
	\quad
	\subfloat[Predicted\label{fig:WESval2}]{
		\includegraphics[scale=0.48]{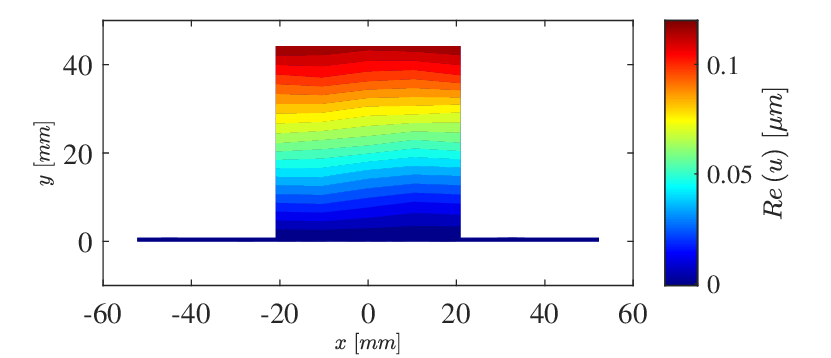}
	}\\
	\subfloat[Ground truth\label{fig:WESval3}]{
		\includegraphics[scale=0.48]{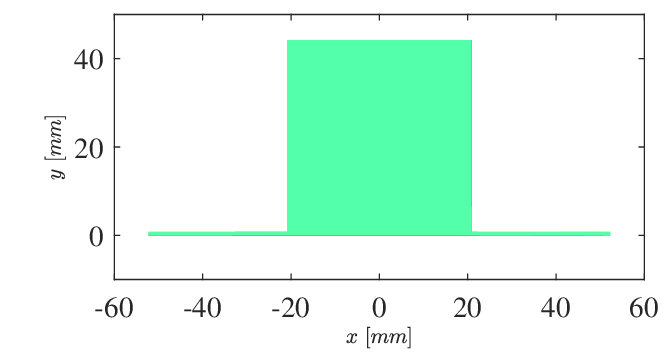}
	}
	\quad
	\subfloat[Predicted\label{fig:WESval4}]{
		\includegraphics[scale=0.48]{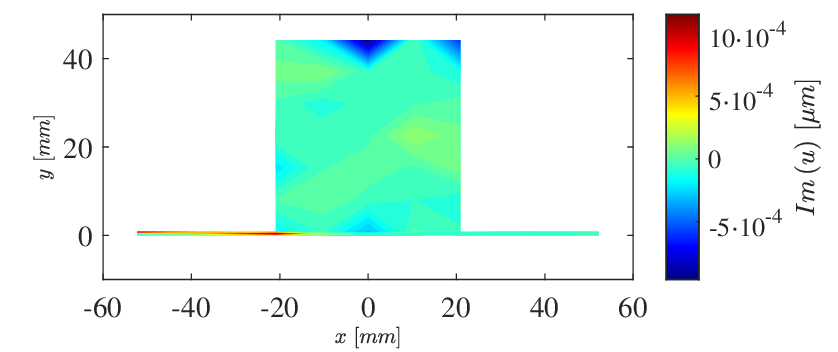}
	}\\
	\subfloat[Ground truth\label{fig:WESval5}]{
		\includegraphics[scale=0.48]{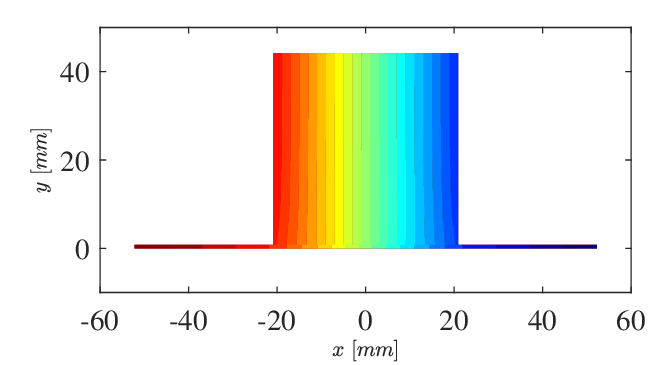}
	}
	\quad
	\subfloat[Predicted\label{fig:WESval6}]{
		\includegraphics[scale=0.48]{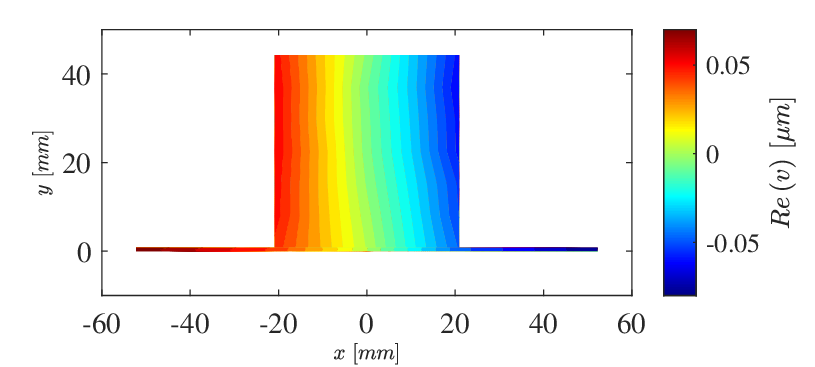}
	}\\
	\subfloat[Ground truth\label{fig:WESval7}]{
		\includegraphics[scale=0.48]{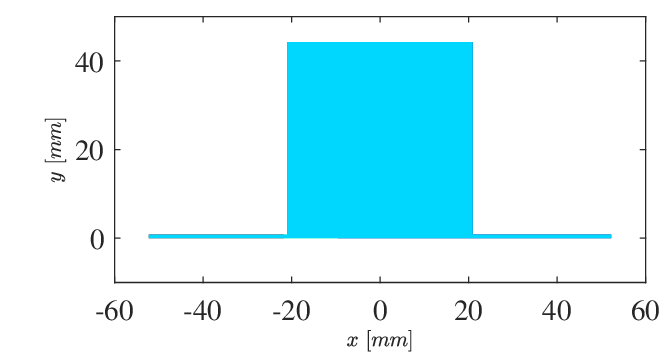}
	}
	\quad
	\subfloat[Predicted\label{fig:WESval8}]{
		\includegraphics[scale=0.48]{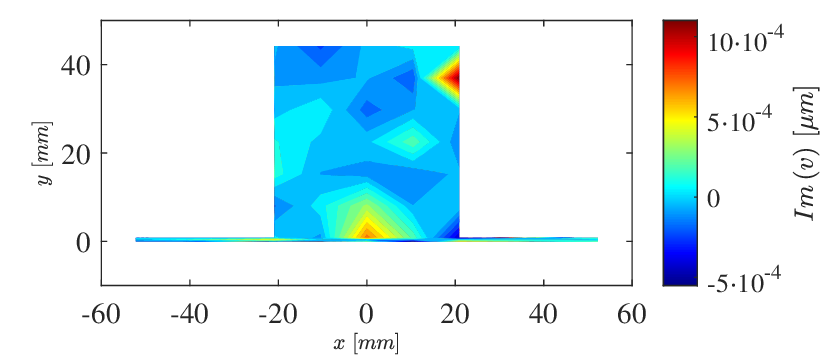}
	}
	\caption[Validation WES model - low frequency]{Graphic validation of the WES model for each component of the eigenvector identified by $n=1$ and $\kappa=-30.1\,\frac{rad}{m}$. The real (a, b) and imaginary (c, d) components of the horizontal displacement field, and the real (e, f) and imaginary (g, h) components of the vertical displacement field are shown}
	\label{fig:WESval}
\end{figure}

\begin{figure}[h!]
	\centering
	\subfloat[Ground truth\label{fig:WESval1_2}]{
		\includegraphics[scale=0.48]{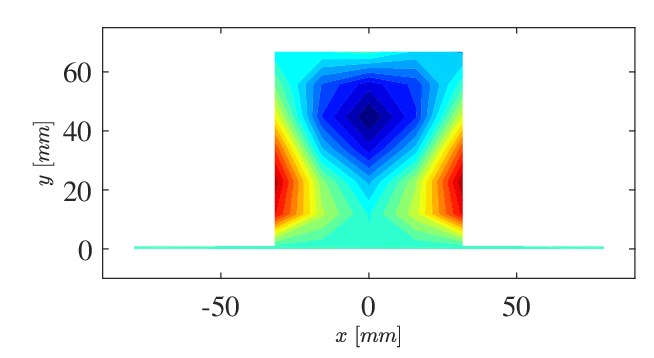}
	}
	\quad
	\subfloat[Predicted\label{fig:WESval2_2}]{
		\includegraphics[scale=0.48]{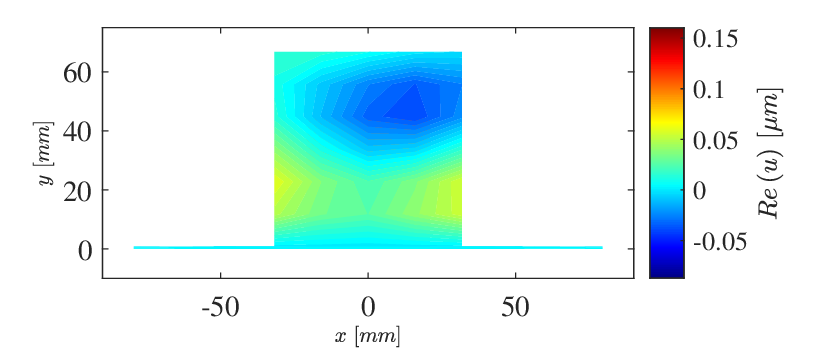}
	}\\
	\subfloat[Ground truth\label{fig:WESval3_2}]{
		\includegraphics[scale=0.48]{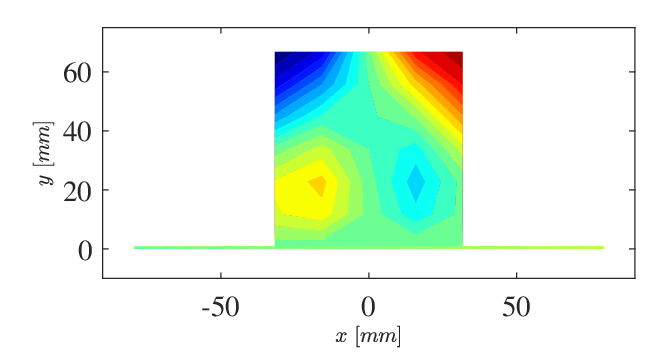}
	}
	\quad
	\subfloat[Predicted\label{fig:WESval4_2}]{
		\includegraphics[scale=0.48]{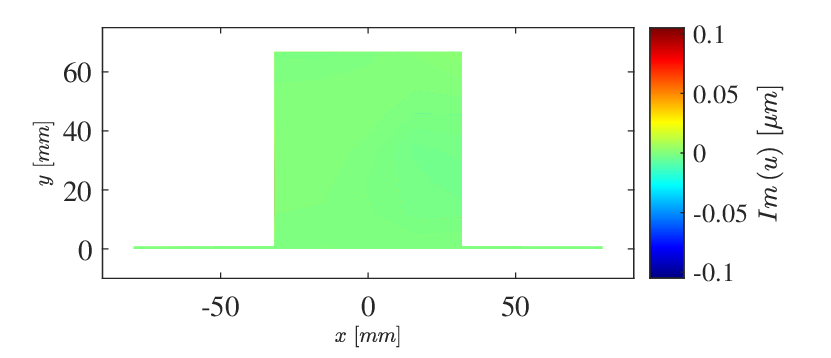}
	}\\
	\subfloat[Ground truth\label{fig:WESval5_2}]{
		\includegraphics[scale=0.48]{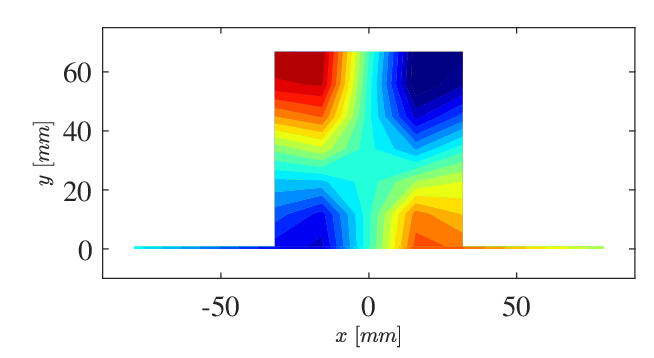}
	}
	\quad
	\subfloat[Predicted\label{fig:WESval6_2}]{
		\includegraphics[scale=0.48]{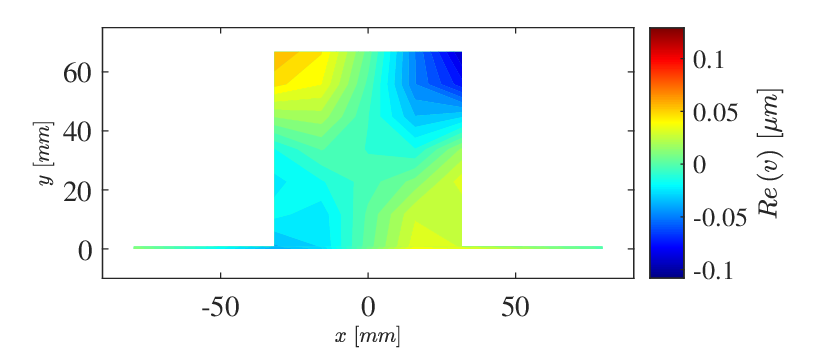}
	}\\
	\subfloat[Ground truth\label{fig:WESval7_2}]{
		\includegraphics[scale=0.48]{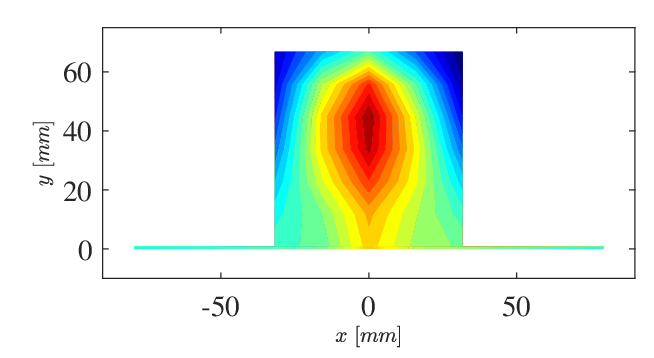}
	}
	\quad
	\subfloat[Predicted\label{fig:WESval8_2}]{
		\includegraphics[scale=0.48]{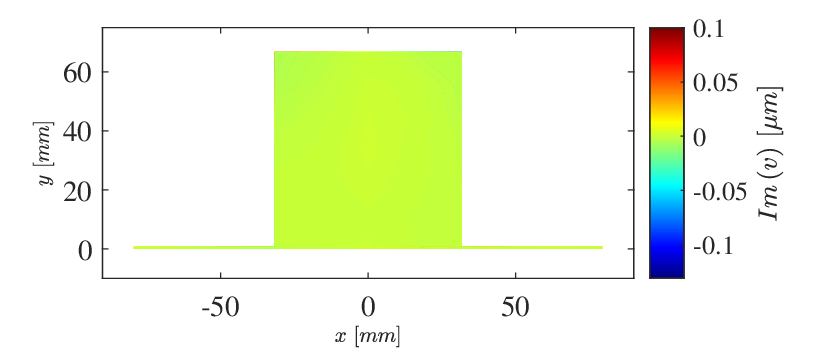}
	}
	\caption[Validation WES model - high frequency]{Graphic validation of the WES model for each component of the eigenvector identified by $n=7$ and $\kappa=-0.989\,\frac{rad}{m}$. These are the horizontal displacement field, (a,b) real and (c,d) imaginary part, and the vertical one, (e,f) real and (g,h) imaginary part.}
	\label{fig:WESval_2}
\end{figure}

The same graphic validation can also be performed for the WES model. The geometry taken in by the model and the material properties are constructed with the ground-truth parameters, and then the predicted eigenvectors are compared with the FE solution. Figs. \ref{fig:WESval} and \ref{fig:WESval_2} represent an example of a low-frequency and a high-frequency eigenvector, respectively.
By comparing Figs.~\ref{fig:WESval} and \ref{fig:WESval_2}, it can be observed that low-frequency eigenvectors are reproduced with greater accuracy than their high-frequency counterparts. Observing the real parts of the displacements in Fig.~\ref{fig:WESval_2}, the field appears down-scaled relative to the ground truth of the FE. Although scaled eigenvectors remain valid solutions to the eigenvalue problem mathematically, the WES is trained to reconstruct them based on labels that maintain a fixed scale. Therefore, the spectral bias introduced by the IEPS also has a non-negligible effect on the WES, leading to inaccurate results above the 3\textsuperscript{rd} eigenfrequency.

\begin{figure}[h!]
	\centering
	\subfloat[\label{fig:SimParams}]{
		\includegraphics[scale=0.49]{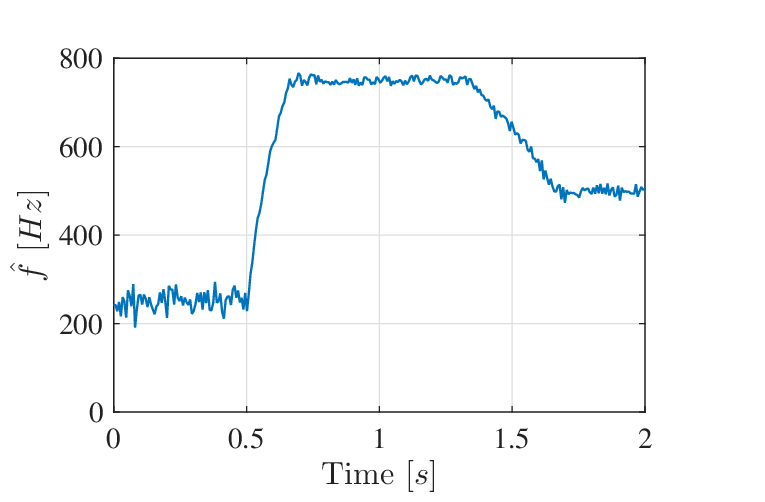}
	}
	\quad
	\subfloat[\label{fig:SimTarget}]{
		\includegraphics[scale=0.49]{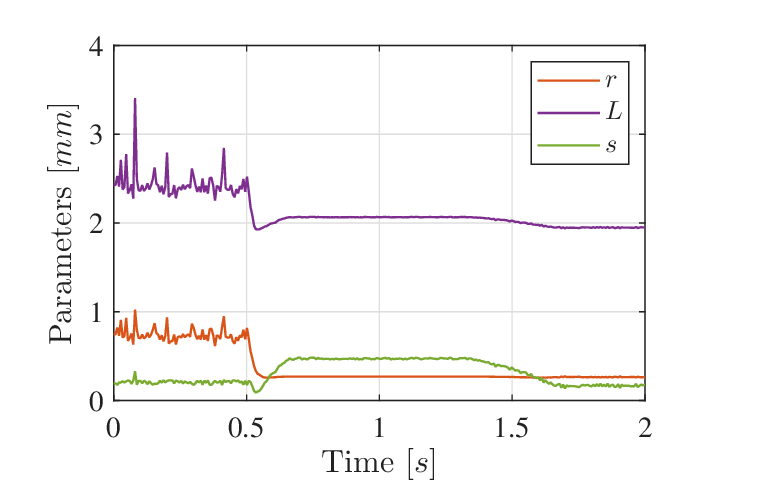}
	}
	\caption[Simulation of SICE4 application]{Simulation of application of the SICE4 algorithm to a real world time series with (a) being the input of target frequencies and (b) the corresponding estimated structural parameters}
	\label{fig:Sim}
\end{figure}

After validation, DeepF-fNet was then tested as part of the SICE4 algorithm. As already mentioned in \autoref{subsec:SICE4}, the SICE4 algorithm is necessary to efficiently estimate structural parameters starting from the user-defined frequency to dampen. The first test of the algorithm was a simulated time series of target frequencies to stop, which is important to evaluate the ability to dynamically adjust the parameters. The simulated signal passes from $250\,Hz$ to $750\,Hz$, and then back to $500\,Hz$ with transients having different speeds. The fictitious time window is $2\,s$ with a sampling frequency of $150\,Hz$, resulting in $300$ target frequencies. Gaussian noise (mean: $0\,Hz$, standard deviation: $100\,Hz$) was added to render the signal more realistic. The target frequencies to stop and the corresponding optimal parameters estimated by the SICE4 algorithm are represented in Fig.~\ref{fig:Sim}.
The sensitivity of each parameter to achieving varied targets can be studied by examining which signal demonstrates the highest noise level among the three parameters at each frequency plateau. The small variations at low frequencies are driven mainly by $L$, while the same variations at high frequencies are driven by $s$. This effect can be interpreted with knowledge of the two different mechanisms of bandgap formation \cite{BARAVELLI20136562}:
\begin{itemize}
	\item \textbf{Low frequencies}: the resonance of the ligaments of the ALR lattice is dominant, and because the ligaments themselves can be approximated as encastered beams, the length of the beam $L$ is notoriously the parameter most influencing the eigenfrequencies.
	\item \textbf{High frequencies}: the chiral deformation mechanism is dominant, which means that the circles of the lattice rotate antisymmetrically. This rotation stretches the ligaments, which therefore vibrate longitudinally; hence, the natural frequency is governed by the cross-sectional area. Since the out-of-plane depth is constant, the remaining dominant parameter is the thickness of the ligaments $s$.
\end{itemize}

\begin{figure}[h!]
	\centering
	\subfloat[SICE4\label{fig:SICE4_DC}]{
		\includegraphics[scale=0.49]{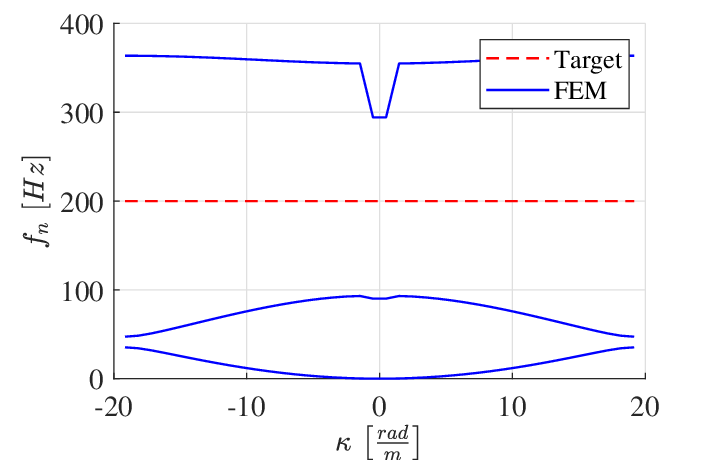}
	}
	\quad
	\subfloat[SICE4\label{fig:SICE4_LRM}]{
		\raisebox{6mm}{\includegraphics[scale=0.17]{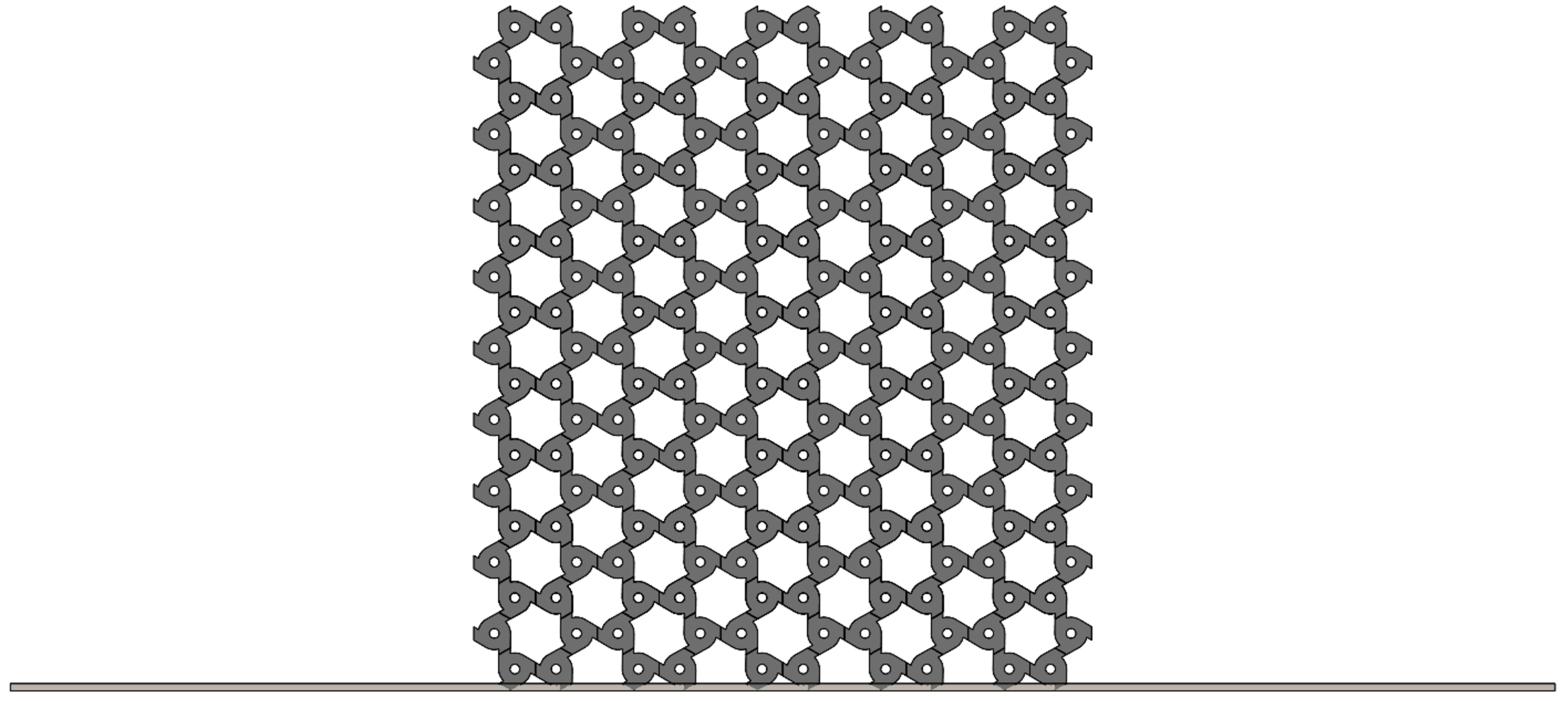}}
	}\\
	\subfloat[GA\label{fig:GA_DC}]{
		\includegraphics[scale=0.49]{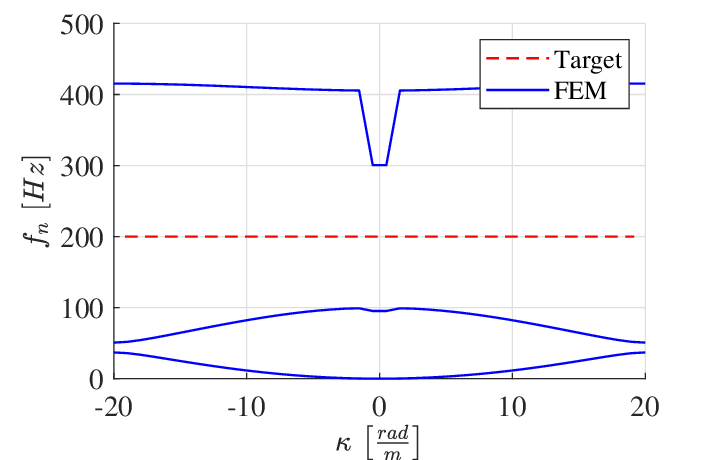}
	}
	\quad
	\subfloat[GA\label{fig:GA_LRM}]{
		\raisebox{6mm}{\includegraphics[scale=0.17]{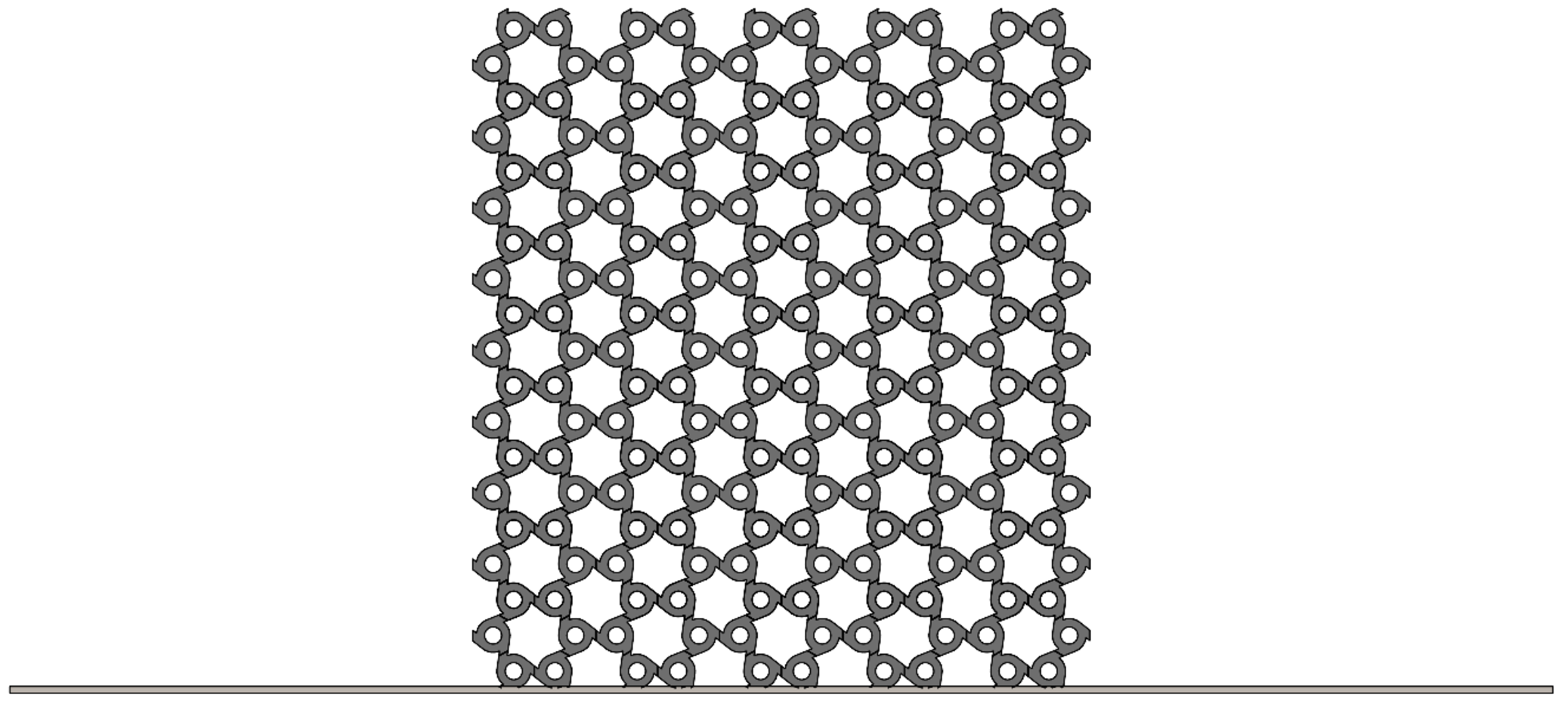}}
	}
	\caption[SICE4 versus GA at $200\,Hz$]{Application of the SICE4 and GA algorithms to the single target frequency of $200\,Hz$, showing (a,c) a zoom on the first three eigenfrequencies of the dispersion curves and (b,d) the LRM unit cell with the optimal parameters}
	\label{fig:SICE4}
\end{figure}

The second test involved comparing the performance of SICE4 with that of a genetic algorithm (GA) to estimate the optimal parameters of the unit cells to filter out a single target frequency of $200\,Hz$. The GA used in this work is based on the work of Sayin \cite{GA}. This comparison is crucial because it highlights the advantages of the proposed method relative to the current state-of-the-art. The optimal parameters identified by the algorithms were used to generate the corresponding dispersion curves using FE simulations. The analysis presented in this work focuses on the first three dispersion curves, that is, it focuses on the first bandgap (Figs.~\ref{fig:SICE4_DC} and \ref{fig:GA_DC}).
\begin{table}[h!]
	\centering 
	\begin{tabular}{ c  c  c }
		\hline
		& \textbf{GA} & \textbf{SICE4} \\
		\hline \hline
		\textbf{r} $\bm{\left[mm\right]}$ & 0.968 & 1.12 \\
		\textbf{L} $\bm{\left[mm\right]}$ & 3.72 & 3.76 \\
		\textbf{s} $\bm{\left[mm\right]}$& 1.46 & 1.23 \\
		\textbf{CPU time} $\bm{\left[s\right]}$& 7690 & 0.0157\\
		\hline
	\end{tabular}
	\\[10pt]
	\caption{Comparison between the GA and the SICE4 results}
	\label{tComp}
\end{table}
As can be observed in Tab.~\ref{tComp}, the optimal parameters predicted by both methods are almost identical, with the resulting bandgaps reflecting this similarity (Figs.~\ref{fig:SICE4_DC} and \ref{fig:GA_DC}).
However, in terms of computational speed, the SICE4 algorithm far outperforms GA, achieving a performance $5\cdot 10^{5}$ times faster than the other algorithm. This significant advantage is primarily due to the fact that the most time-consuming phase, i.e., the training of the DeepF-fNet, is performed offline and therefore does not affect the computation time during optimization.

%% file: sections/conclusions.tex
\section{Conclusions}
\label{sec:conclusions}
In this work, we have introduced a new framework, DeepF-fNet, which uses DeepONet and physics-informed neural networks for structural optimization in vibration isolation. By incorporating physical laws into training, DeepF-fNet reduces data requirements and improves accuracy in various scenarios, offering a significant computational advantage. The SICE4 algorithm, developed to complement DeepF-fNet, enables real-time predictions by implementing a prediction-correction approach, delivering results up to 500~000 times faster than traditional methods like genetic algorithms. This computational speed and adaptability suggest strong potential for use in real-time systems, especially in applications demanding rapid optimization, such as automotive semi-active vibration isolation systems.

The framework was validated with a case study on locally resonant metamaterials, demonstrating accuracy in predicting low-frequency dispersion curves and mode shapes, and confirming its superior computational speed compared to classical optimization. Although spectral bias at higher eigenfrequencies presents a limitation, further refinement through Fourier neural operators ~\cite{Fourier,li2021fourierneuraloperatorparametric} is proposed to improve predictions for multi-bandgap targets.

Future improvements will also include dataset expansion and experimental validation. Increasing dataset diversity would strengthen model robustness at different frequency ranges. Experimental validation, such as modal analysis on a sensorized steel plate with ALRs, is recommended to confirm numerical predictions. This validation would assess model accuracy and practical applicability, marking a crucial step in advancing DeepF-fNet and SICE4 for deployment in complex, resource-constrained environments.